\documentclass[5p,times]{elsarticle}

\usepackage{textcomp,gensymb}
\usepackage{multirow}
\usepackage{float}
\usepackage[latin1]{inputenc}   
\usepackage{amsmath} 
\usepackage{epstopdf}           %
\usepackage{flushend}           %
\usepackage{lmodern}
\usepackage[T1]{fontenc}
\usepackage{xcolor}
\usepackage{upgreek}

\usepackage{amssymb}

\usepackage{lineno}

\usepackage[figuresright]{rotating}
\usepackage{color,soul}
\usepackage{siunitx}

\usepackage{todonotes}
\usepackage{subfigure}

\begin{document}
\begin{frontmatter}

\title{Non-destructive 3D characterization of microtextured regions in the bulk of Ti-6Al-4V alloy}

\author[First]{Mads Carlsen\corref{cor1}}
\ead{mads.carlsen@psi.ch}
\author[Second]{Xiaohan Zeng}
\author[Third]{Haixing Fang}
\author[Fourth]{Moritz Frewein} 
\author[Fourth]{Tilman A. Gr\"unewald}
\author[Second]{Joao Quinta da Fonseca}
\author[Fifth,Third]{Wolfgang Ludwig}

\cortext[cor1]{Corresponding Author}

\address[First]{Paul Scherrer Institut, 5232 Villingen, PSI, Switzerland}
\address[Second]{Department of Materials, University of Manchester, Oxford Road, M139PL Manchester, UK}
\address[Third]{European Synchrotron Radiation Facility, 71 Avenue des Martyrs, 38043 Grenoble Cedex 9, France.}
\address[Fourth]{Aix Marseille Univ, CNRS, Centrale Med, Institut Fresnel, 13397 Marseille, France}
\address[Fifth]{Univ Lyon, CNRS, INSA Lyon, Universit\'e Claude Bernard Lyon 1, MATEIS, UMR5510, 69100 Villeurbanne, France}

\date{\today}

\begin{abstract}
In this study we present spatially resolved texture and orientation maps from a cube-shaped sample of Ti-6Al-4V alloy, reconstructed by means of texture tomography (TT). Unlike grain resolved 3DXRD techniques which require "spotty" diffraction patterns, TT can reconstruct local (voxelized) orientation distribution function (ODFs) from continuous diffraction patterns collected in a 3D scanning procedure. Reconstructions of the same sample, scanned in two different settings and subsequent EBSD analysis parallel to one of the sample faces show excellent mutual agreement and validate the single-axis data collection and reconstruction procedure for this class of materials. The reconstruction reveals the presence and the 3D shape of several micro-textured regions, showing a sharp unimodal texture of the $\alpha$-phase with a clear link of the orientation and spatial alignment of these zones to the rolling, transverse and normal directions of the rolled material.

\end{abstract}


\end{frontmatter}
\begin{sloppypar}
Ti-6Al-4V (Ti64) is the most commonly used two-phase $(\alpha + \beta)$ titanium alloy in aerospace applications due to its high strength-to-weight ratio, corrosion resistance, and high strength at moderate temperatures. Depending on the thermo-mechanical processing route, Ti64 may develop micro-textured regions (MTRs) exhibiting preferential crystallographic orientation of the hexagonal $\alpha$ phase, with dimensions ranging from 0.1 mm to cm \cite{zeng2025effect}. When subject to mechanical loads, the presence of MTRs gives rise to heterogeneities in the local elastic and plastic response of the material \cite{Hijazi2023, Lunt2017, Cappola2021} and the resulting stress concentrations are known to be detrimental to the dwell fatigue life of components like discs in jet engines \cite{Liu2021}. These MTRs have a complex microstructure. Several non-destructive testing (NDT) methods exist to reveal the presence of MTR's intersecting the material surface: electron back-scatter diffraction (EBSD) \cite{Germain2005, Davis2022} and polarized light microscopy (PLM) \cite{Bhme2018, Jin2020, Hijazi2022} are routinely deployed for this purpose. When it comes to NDT of macrozones in the bulk of the material, different variants of ultrasonic inspection techniques are actively researched \cite{Moreau2013, Lan2018, Baelde2018, Yeoh2023}. Although they are capable to detect the presence of MTRs, so far these techniques can not reliably characterize the orientation and actual shape of sub-surface MTRs. 
Neutron and synchrotron X-ray diffraction techniques allow to reconstruct the volume average texture, more precisely the orientation distribution function (ODF) by measuring 2D diffraction patterns, collected at a few different sample orientations \cite{Daniel2023}. When combined with a 2D scanning approach, the local texture, averaged over the sample thickness, can be mapped. 
Here we present an extension of the 2D texture mapping approach into 3D, using the concepts of texture tomography (TT)\cite{frewein_2024}. 

\par
A cube shaped sample with 2 mm edge length was extracted via electrical discharge machining from a $20\,\si{mm}$ thick sheet of Ti64 with  the edges of the cube aligned with the rolling direction (RD), transverse direction (TD), and normal direction (ND). The material had been hot rolled at $865\,\si{\degree C}$ to 50\% reduction as described in \cite{zeng2025effect}. 

\par
The X-ray measurements were performed on the 3DXRD endstation of the materials science beamline ID11 at the European Synchrotron Radiation Facility\cite{Wright2020} using an X-ray energy of 70keV. The cube shaped sample was mounted on a magnetic support frame, allowing for $120\si{\degree}$ rotations around an axis inclined by $35\si{\degree}$ with respect to the incoming X-ray beam. During the first acquisition, the cube face with normal parallel to the ND of the material was oriented along the (Z) rotation axis of the instrument and the X-ray beam was parallel to RD at $0\si{\degree}$ diffractometer rotation angle $\omega$. For each sample rotation position, TT projections were collected as 2D raster scans of the sample translation stages, moving in the laboratory Y and Z direction, orthogonal to the incoming beam. For each sample position, a 2D diffraction image was collected on a Dectris Eiger 4M detector, positioned at $324\,\si{mm}$ from the rotation axis and operating at a frame rate of 100 Hz. By scanning the vertical sample translation axis in continuous mode, the acquisition time for the collection of one of these 115 (Y) x 88 (Z) pixel projection images, including motion and software overheads, was 330 seconds.  A total of 120 equi-spaced projections were collected over a rotation range of $180\si{\degree}$. A second series of TT projections were acquired in the rotated sample position (see figure~S1.) A full tomographic measurement of the sample took 11 hours of which 3.4 hours were measurement and the rest overheads due to scanning. The ongoing upgrade of the ID11 diffractometer will allow for a double-continuous layer-by-layer scanning procedure (e.g. continuous rotation and translation perpendicular to the rotation axis), thereby minimizing dead-time for motor acceleration. At an instrument rotation speed of 1 Hz, the scan time for the current sample reduces to about 1.5 hours. 


\par
The recorded diffraction patterns are first azimuthally regrouped into radial bins (12 for the $\alpha$-phase and 2 for the $\beta$-phase), each covering the full width of a separate diffraction peak and 180 azimuthal ($\eta$) bins of $2\si{\degree}$ width which reduces the size of the dataset down to around 60GB for each 3D scan. The data is then corrected for various experimental errors and normalized by X-ray transmission and structure factor. The regrouped data is corrected for absorption using the known sample-shape and absorption coefficient using the "small-angle approximation"\cite{Carlsen_2025_absorp}. The structure factors are estimated from the orientationally and spatially averaged diffraction profile. The result is a five dimensional TT dataset ($\omega, \mathrm{Scan}\, Y, \mathrm{Scan}\, Z, \eta, 2\theta$). 

\par
In TT, the sample is modeled by a regular 3D array with a voxel-size matching the raster-scan step size where each voxel contains the coefficients of a series expansion of an ODF. When the voxel size is not much larger that the typical grain size of the sample, each voxel only contains a small number of grains. Therefore, the ODF in any individual voxel is expected to only be non-zero in the neighborhood of the orientations of these grains. In this case, a series expansion in terms of radial basis functions (RBF) is well-suited as the coefficients of the converged solution will be highly sparse which can be used to effectively constrain the tomographic reconstruction problem\cite{Carlsen_2025}.

\par
Reconstruction is performed by fitting the texture model to the azimuthally regrouped diffraction data via an iterative gradient-descent algorithm. In typical parallel-beam geometry where a single rotation axis perpendicular to the incoming X-ray beam is used, the model can be fitted in a slice-by-slice fashion, easing the computational requirements compared to simultaneous 3D reconstructions which are necessary with more complicated measurement geometries that have previously been used in TT\cite{frewein_2024, Carlsen_2025}. The reconstructed coefficients provide a good fit to the data ($R^2 = 0.95$) with $18.5\%$ nonzero coefficients leading to 1.5 billion nonzero model parameters and 1.0 billion data points.

\par
For cross-validation of the TT reconstruction, the ND-TD sample surface was prepared by grinding to 4000 grit, followed by a 5-minute polish using a 5:1 mixture of colloidal silica and hydrogen peroxide. Electron backscatter diffraction (EBSD) was performed on this ND-TD plane using a ThermoFisher Apreo scanning electron microscope equipped with an Oxford Instruments Symmetry EBSD detector. The accelerating voltage and beam current were set to 20 kV and 26 nA, respectively.

\begin{figure}[t]
    \centering
    \resizebox{1.0\columnwidth}{!}
    {\includegraphics{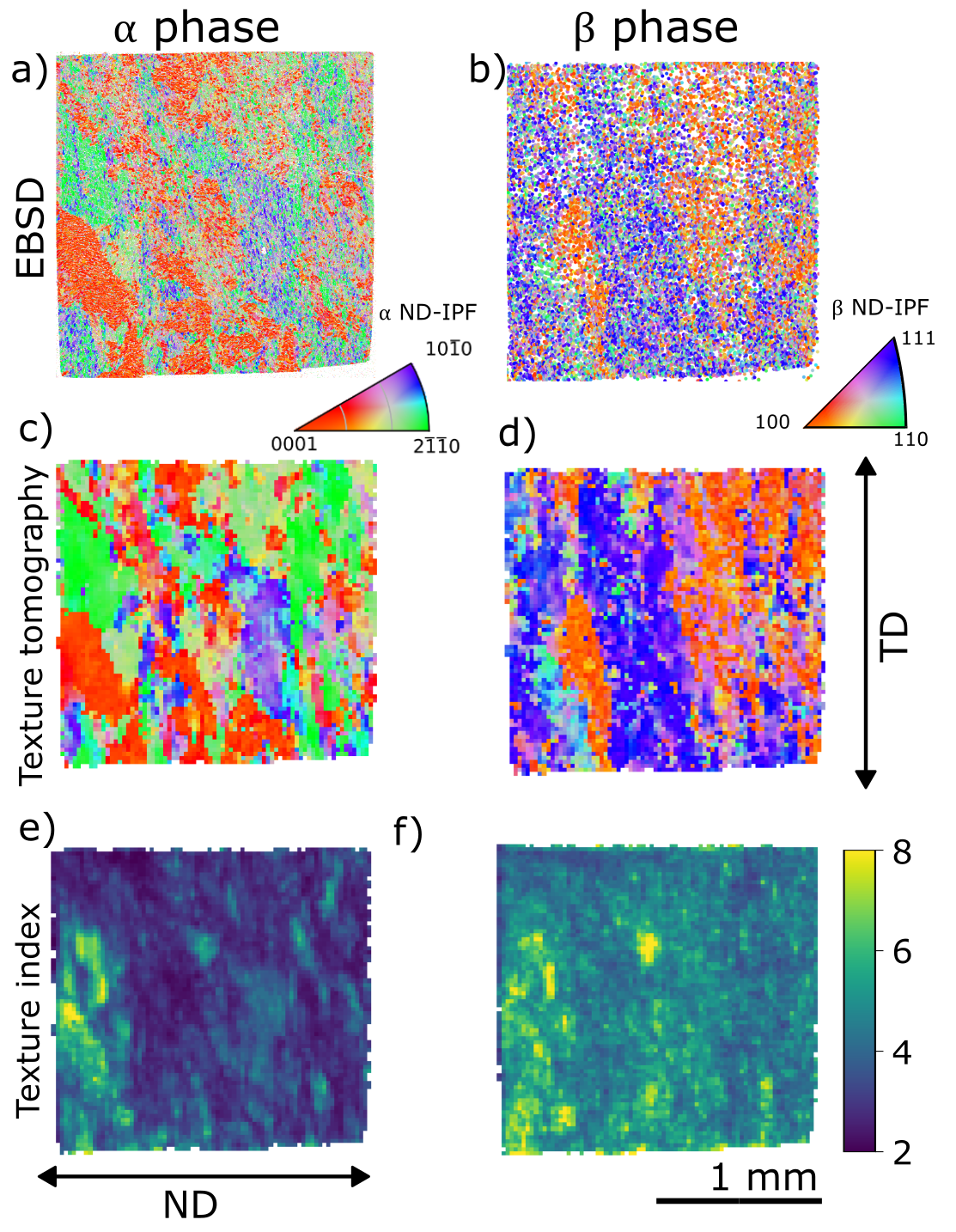}}
    \vspace{-5mm}
    \caption{ND IPF maps of a TD-ND plane close to the front surface of the sample cube. a,c) show the orientation of the $\alpha$-phase and b,d) show the orientation of the $\beta$-phase as found by a,b) EBSD and c,d TT respectively. As only a small fraction (2.7\%) of the points were indexed as $\beta$ phase by EBSD, the indexed points are plotted as enlarged circles for easier visualization. e,f) shows the texture index computed from the TT reconstructions. }
    \label{fig:compare_with_EBSD}
    \vspace{-5 mm}
\end{figure}

This comparison is shown in Fig. \ref{fig:compare_with_EBSD}. Since the individual grains are not resolved by the tomographic reconstruction and the EBSD measurement is only sensitive to a thin surface layer, much smaller than the voxel-size of the tomographic reconstruction, it is difficult to perform a quantitative comparison between the two, but visually there is a strong correlation, especially for the highly textured regions.

\begin{figure}[t]
    \centering
    \resizebox{1.0\columnwidth}{!}
    {\includegraphics{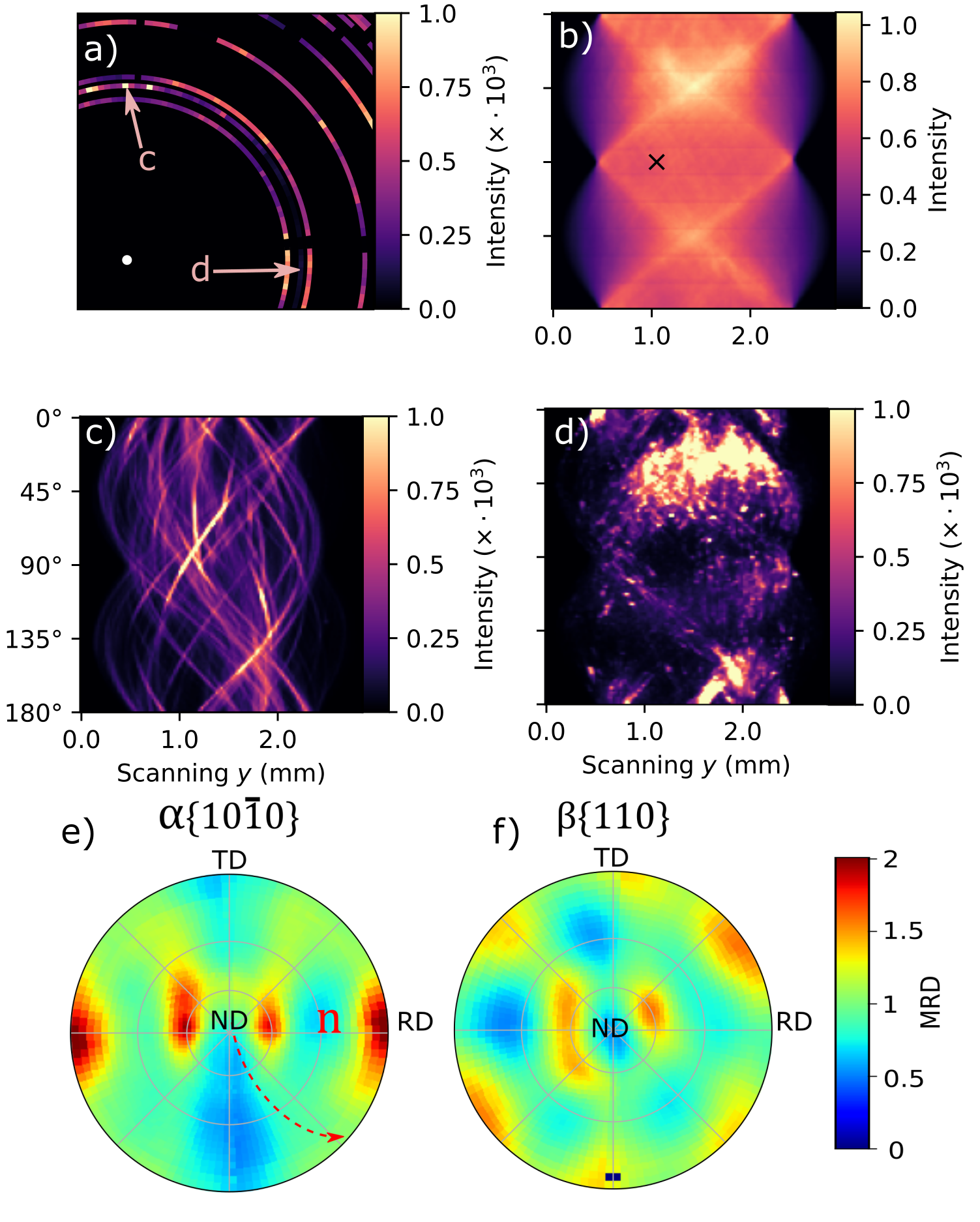}}
    \vspace{-8mm}
    \caption{Examples of measured scanning-diffraction data. a) Normalized scattering pattern from a single sample position after azimuthal regrouping. b) Normalized intensity summed over $hk\ell$-orders and azimuthal angle at a fixed sample height as a function of tomographic angle $\omega$ and scanning $y$-position. c,d) Normalized intensities of a single detector segment, both from the $0002$ diffraction ring at azimuthal angles marked in a). Experimental sample-averaged pole figures of the e) $\alpha$-phase $\{10\overline{1}0\}$ and f) $\beta$-phase $\{110\}$ reflection. The axis of the manual rotation is marked by $\mathbf{n}$ and the sense and magnitude is indicated by a red arrow.}
    \label{fig:scattering_sinograms}
    \vspace{-5 mm}
\end{figure}

Building upon this validation of the methodology, we now analyse texture at different scales, starting from sample average over MTR-average down to single MTRs and individual voxels. The sample has weak texture on average with a $\{0002\}$ preferentially oriented along TD and $\{10\overline{1}0\}$ oriented along RD, in agreement with the macroscopic texture of the material it was extracted from \cite{zeng2025effect}. The sample averaged experimental pole figures (Fig.\ref{fig:scattering_sinograms} e,f) lacks sharp features, as expected from an as-deformed, unrecrystallized microstructure. It furthermore appears that the sample volume is not large enough to reach statistical sample symmetries expected for rolling textures even though it contains an enormous number of grains: ($(2\,\si{mm})^3 / (15\,\si{\micro m})^3 \approx 2.4\cdot 10^6 $). This deviation demonstrates how the local deformation in these materials during rolling is different from the overall deformation, which is a consequence of the strong plastic anisotropy of the hexagonal $\alpha$ phase and the presence of MTRs, i.e. strong microtextures despite a weak macrotexture. Rietveld refinement of the orientationally averaged data found 8\% volume fraction of the $\beta$ phase.


\par
The sample averaged pole figures contain regions of high intensity with a characteristic full width of around $20\si{\degree}$. The spatially resolved diffraction data contains a mix of such broad diffraction features with sharper features down to a few degrees in azimuthal widths (Fig. \ref{fig:scattering_sinograms} a). These sharp features are limited in spatial extent, typically to a single scan point (Fig. \ref{fig:scattering_sinograms} c,d) and originate from individual grains while the broader features can have large spatial extent - up to almost the full width of the sample, exemplified by the bright feature in  Fig. \ref{fig:scattering_sinograms}d) near $\omega = 45\si{\degree}$ and originate from highly textured regions where the individual grains cannot be distinguished in the diffraction patterns. By inspecting the three dimensional data (see video in supplementary material) it is apparent that these originate from the plate-like MTRs.
 
\par
\begin{figure*}[t]
    \centering
    \resizebox{1.5\columnwidth}{!}
    {\includegraphics{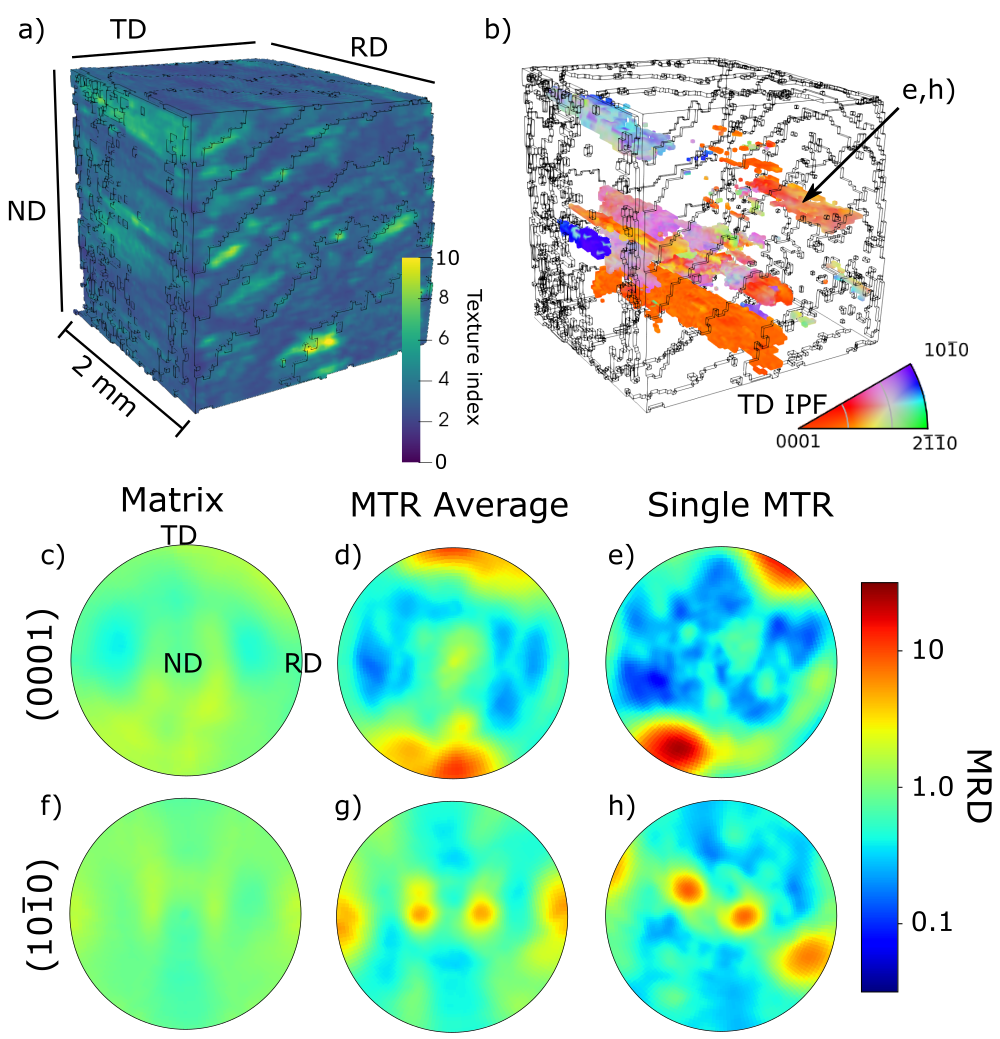}}
    \vspace{-1mm}
    \caption{a) 3D rendering of the reconstructed cube colored by texture index. b) only regions of high texture index coloured by the inverse pole figure color of the mean orientation. c-e) $(0001)$ pole figures of c) regions with low texture index, d) regions with high texture index, and e) a single textured region. f-h) $(10\overline{1}0)$ pole figure of the same three textures as c-e). }
    \label{fig:reconstruction_plot}
    \vspace{-5 mm}
\end{figure*}

The TT reconstruction reveals both the shape and the crystallographic orientation of the MTRs. A 3D map of the voxel-by-voxel texture index is visualized in Fig. \ref{fig:reconstruction_plot}a), plotted on three surfaces of the sample. The MTRs show up as distinct regions of high texture index in a matrix of lower texture index. 

\par
On average, the MTRs have a texture with $(0001)$ parallel to TD, $(2\overline{1}\overline{1}0)\,||\,\mathrm{ND}$, and $(01\overline{1}0)\,||\,\mathrm{RD}$ but display a large scatter around this orientation on the order of $30\si{\degree}$. The individual MTRs are significantly more ordered with a FWHM of around $20\si{\degree}$ (Fig. \ref{fig:reconstruction_plot} e,h). The MTRs explain most of the macroscopic texture, which can be seen by comparing the pole figures of the MTRs (Fig. \ref{fig:reconstruction_plot} d,g) with the experimental sample-averaged pole figures (Fig. \ref{fig:scattering_sinograms} e) and by the almost random texture of the matrix (Fig. \ref{fig:reconstruction_plot} c,f). 

\par


The cubic $\beta$-phase displays broader diffraction peaks than the $\alpha$ phase. Only two non-parallel diffraction rings, (110) and (200), were used to reconstruct the texture of the $\beta$-phase material due to an incidental overlap of the $\beta \,(211)$ with the much more intense $\alpha \,(10\overline{1}3)$ and the low volume fraction which meant that rings at higher scattering angle than (310) were barely above noise level. Therefore the $\beta$ phase reconstruction is of lower resolution, as confirmed by Fourier analysis of the validation experiment (Fig. S3).

\begin{figure}[t]
    \centering
    \resizebox{0.99\columnwidth}{!}
    {\includegraphics{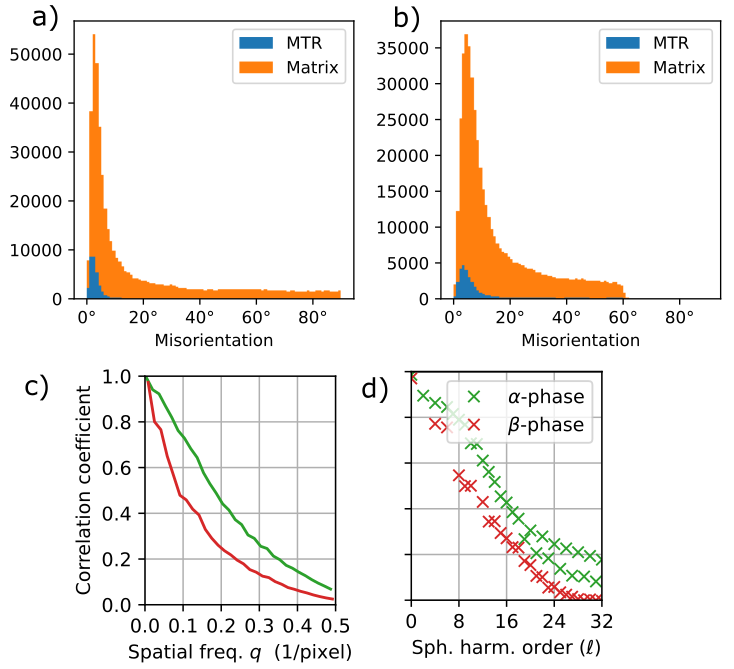}}
    \vspace{-1mm}
    \caption{ Stacked-histograms of voxel-by-voxel orientation error of two independent reconstructions after back-interpolation to a common coordinate system for the a) $\alpha$ and b) $\beta$ phases, respectively. c) Pearson correlation coefficient as a function of spatial resolution and d) as a function of angular resolution. Definition of correlation is provided in the supplementary information.}
    \label{fig:mean_orientation_histograms}
    \vspace{-5 mm}
\end{figure}

To further validate the reconstruction approach, we measure two independent tomographic datasets where the sample has been manually rotated, such that the tomographic axes are orthogonal (Fig. S1). From each data set an independent reconstruction is created. To compare the two reconstructions, we first rotate the second into the coordinate system of the first using tri-linear interpolation. Then the computed main-orientation is compared voxel for voxel. Figure \ref{fig:mean_orientation_histograms}\,a,b) shows histograms of the per-voxel misorientation magnitude. The histograms display a narrow peak with mode at $2\si{\degree}-3\si{\degree}$ for $\alpha$ and $4\si{\degree}-5\si{\degree}$ for $\beta$ which contains about half of the voxels and a broad distribution containing the remaining voxels. Looking only at the MTR, where the mean orientation is well-defined, the mean orientation errors are $4.1\si{\degree}$ and $10.0\si{\degree}$ for $\alpha$- and $\beta$ phases respectively.

For a more quantitative measure of the similarity that takes into account the full ODF, and not just the main orientation, we compute a orientationally averaged correlation as a function of spatial frequency (Fig. \ref{fig:mean_orientation_histograms}c) and a spatially averaged correlation as a function of orientation resolution (Fig. \ref{fig:mean_orientation_histograms}d). For the $\alpha$ phase, the two reconstructions remain well correlated up to an order of around $\ell = 16$ which corresponds to a half-pitch resolution of about $10\si{\degree}$. The $\beta$ phase reconstructions are less correlated showing a steep drop off directly from zero-frequency, rather than displaying a plateau as observed for the $\alpha$ phase. This is likely due to the smaller number of $hk\ell$ orders used in the reconstruction and the worse signal-to-noise ratio due to the low volume fraction.

The cross-validation of results obtained with the $120\si{\degree}$ rotated sample and the good agreement with EBSD measurements, validates the use of a single rotation axis. This allows a 2D slice-by-slice reconstruction procedure which greatly reduces the memory footprint of the reconstruction problem. With the current software implementation, lateral scan dimensions (perpendicular to the rotation axis) of a few hundred positions can be handled. Keeping a step size of $25\,\si{\micro m}$ (suitable for rolled Ti64), lateral sample dimensions up to 10 mm could be handled by adapting the X-ray energy. These indicative limits strongly depend on the X-ray attenuation of the material and have to take into account the safe energy operation range of the detector hardware and size of the micro-textured regions to be characterized. 
   

The current approach can handle large intragranular orientation spread, as encountered in materials which have undergone high levels of plastic deformation, or which contains grains of very different sizes and phases as encountered in additive manufacturing techniques but also in advanced steels and other titanium alloys.
TT shares an identical experimental setup with grain resolved diffraction techniques like scanning-3DXRD\cite{Bonin_2014, Hayashi2015, Henningsson2024} and can be readily combined with phase-contrast tomography. This in turn allows for multi-modal and time resolved characterization of phase fraction, crystallographic texture and microstructure across a wide range of length-scales (100 nm to 1 mm voxels) and opens access to microstructural conditions, previously not accessible by indexing based 3DXRD methods, requiring spotty diffraction patterns. This approach also opens up the possibility of studying, in-situ and through correlative tomography, for example, the deformation and damage behaviour of microstructures in actual engineering materials, overcoming the need to use model materials currently imposed by other techniques. 


In conclusion,  the present study demonstrates the feasibility of 3D spatially resolved texture mapping in an industrially relevant alloy system. A cube shaped sample extracted from a hot-rolled sheet of Ti6Al4V alloy has been measured by texture tomography in two different geometric settings. After rotation to a common reference frame, the two independently reconstructed texture maps are in good mutual agreement, as demonstrated by a correlation analysis combining real-space and orientation space information. The reconstructed (main) orientation map has been further validated using EBSD, acquired close to one of the original sample surfaces. Given the consistent mutual agreement between the different observations we conclude that, for this class of materials, acquisitions using a single tomographic rotation axis are sufficient to reconstruct spatially resolved orientation distribution functions. The spatial and orientation resolution achieved in this study - better than 50 $\micro$m and $10\si{\degree}$ respectively, are sufficient to characterise the 3D shape, orientation and texture index of micro-textured regions (MTRs) in this industrially relevant alloy system. The reconstruction reveals the presence and the 3D shape of several micro-textured regions, showing a sharp unimodal texture of the $\alpha$-phase with $(0001)$ parallel to TD, $(2\overline{1}\overline{1}0)||\mathrm{ND}$, and $(01\overline{1}0)||\mathrm{RD}$ on average, but a large scatter from region to region. The $\beta$ phase also displays a preferential orientation in these regions but exhibits a wider orientation spread and no clear orientation relationship to the $\alpha$ phase.

\section*{CRediT authorship contribution statement}
\textbf{Mads Carlsen} Investigation, Methodology, Software, Formal analysis, Validation, Writing - Original Draft;
\textbf{Xiaohan Zeng} Investigation, Formal analysis, Validation, Writing - Review and Editing;
\textbf{Haixing Fang} Investigation, Formal analysis, Writing - Review and Editing;
\textbf{Moritz Frewein} Investigation, Methodology, Writing - Review and Editing
\textbf{Tilman Grunewald} Methodology, Resources, Writing - Review and Editing
\textbf{Joao Quinta da Fonseca} Resources, Writing - Review and Editing
\textbf{Wolfgang Ludwig} Conceptualization, Formal analysis, Investigation, Writing - Original Draft;

We acknowledge the ESRF for the provision of beamtime under proposal MI1497 at ID11 and we thank Dr. Jonathan Wright for his support of this work and Dr. Steve Gaudez for assistance with Rietveld refinement.

\bibliography{ref_scripta}

\end{sloppypar}
\end{document}